\documentclass[11pt,twoside,letterpaper]{article} 
\usepackage{times,fancyhdr}
\usepackage[dvips]{graphicx}

\usepackage{amsmath}
\usepackage{amsfonts}


\makeatletter
\def\cleardoublepage{\clearpage\if@twoside \ifodd\c@page\else%
    \hbox{}%
    \thispagestyle{empty}%
    \newpage%
    \if@twocolumn\hbox{}\newpage\fi\fi\fi}
\makeatother

\begin{document}
\title{
\bfseries\scshape on operators generated by density matrix}
\author{\bfseries\itshape V. I. Gerasimenko\thanks{E-mail address: gerasym@imath.kiev.ua}\\
Institute of mathematics of the NAS of Ukraine\\
Kyiv, Ukraine}

\maketitle

\renewcommand{\headrulewidth}{0pt}


\begin{abstract}
\noindent In this survey the possible approaches to the description of the evolution
of states of quantum many-particle systems by means of the possible modifications of
the density operator which kernel known as density matrix are considered. In addition,
an approach to the description of the evolution of states by means of the state of a
typical particle of a quantum system of many particles is discussed, or in other words,
the foundations of describing the evolution of states by kinetic equations are considered.
\end{abstract}

\vspace{.1in}


\vspace{.08in}
                             von Neumann equation, von Neumann hierarchy, BBGKY hierarchy, kinetic equation.

\tableofcontents


\section{Introduction}
The paper deals with the mathematical problems of describing the evolution of states of
quantum many-particle systems by means of operators generated by the density operator
which kernel is known as a density matrix.

As known, a quantum system is described in terms of such notions as an observable and a
state. The functional for the mean value of observables determines a duality between
observables and states. In a consequence of this there exist two approaches to the
description of the evolution of a quantum system of finitely many particles, namely,
in terms of observables that are governed by the Heisenberg equation, or in terms of
states governed by the von Neumann equation for the density operator, respectively
\cite{vonN},\cite{DauL},\cite{G12}.

An alternative approach to the description of states of a quantum system of finitely
many particles is given by means of operators determined by the cluster expansions
of the density operator. They are interpreted as correlation operators. The evolution
of such operators is governed by the von Neumann hierarchy \cite{G12},\cite{GerS},\cite{GP11}.

One more approach to describing a state of many-particle systems is to describe a state
by means of a sequence of so-called reduced density operators (marginal density operators)
governed by the BBGKY (Bogolyubov--Born--Green--Kirkwood--Yvon) hierarchy \cite{BogLect}.
An alternative approach to such a description of a state is based on operators determined
by the cluster expansions of reduced density operators. These operators are interpreted
as reduced correlation operators that are governed by the hierarchy of nonlinear evolution
equations \cite{G12}. On a microscopic scale, the macroscopic characteristics of fluctuations
of observables are directly determined by the reduced correlation operators. The mention
approaches are allowed to describe the evolution of states of quantum systems both with a
finite and infinite number of particles, in particular, systems in condensed states \cite{CGP97}.

In addition, an approach to the description of the evolution of states by means of the state
of a typical particle of a quantum system of many particles is discussed, or in other words,
the foundations of describing the evolution of states by kinetic equations are considered \cite{G17IM}.

Hereinafter we denote the $n$-particle Hilbert space which is a tensor product of $n$ Hilbert spaces
$\mathcal{H}$ by the $\mathcal{H}_n=\mathcal{H}^{\otimes n}$ and we use the usual convention that
$\mathcal{H}^{\otimes 0}=\mathbb{C}$. The Fock space over the Hilbert space $\mathcal{H}$ we denote 
by the $\mathcal{F}_{\mathcal{H}}={\bigoplus\limits}_{n=0}^{\infty}\mathcal{H}_{n}$. The self-adjoint
operator $f_{n}$ defined in the space $\mathcal{H}_{n}=\mathcal{H}^{\otimes n}$ will also be denoted
by the following symbol $f_{n}(1,\ldots,n)$.

Let $\mathfrak{L}(\mathcal{H}_{n})$ be the space of bounded operators
$f_{n}\equiv f_{n}(1,\ldots,n)\in\mathfrak{L}(\mathcal{H}_{n})$ equipped with the operator 
norm $\|.\|_{\mathfrak{L}(\mathcal{H}_{n})}$.
Accordingly, let $\mathfrak{L}^{1}(\mathcal{H}_{n})$ be the space of trace class operators 
$f_{n}\equiv f_{n}(1,\ldots,n)\in\mathfrak{L}^{1}(\mathcal{H}_{n})$ equipped with the norm:
$\|f_{n}\|_{\mathfrak{L}^{1}(\mathcal{H}_{n})}=\mathrm{Tr}_{1,\ldots,n}|f_{n}(1,\ldots,n)|,$
where $\mathrm{Tr}_{1,\ldots,n}$ are partial traces over $1,\ldots,n$ particles. Below we 
denote by $\mathfrak{L}^{1}_0(\mathcal{H}_{n})$ the everywhere dense set of finite sequences 
of degenerate operators with infinitely differentiable kernels with compact supports.

\pagestyle{fancy}
\fancyhead{}
\fancyhead[EC]{V. I. Gerasimenko}
\fancyhead[EL,OR]{\thepage}
\fancyhead[OC]{Operators generated by density matrix}
\fancyfoot{}
\renewcommand\headrulewidth{0.5pt}

\section{The density operator}

For generality, we consider a quantum system of non-fixed, i.e. arbitrary, but finite
number of identical (spinless) particles, obeying the Maxwell--Boltzmann statistics, in
the space $\mathbb{R}^{3}$. For this system observables can be described by means of the
sequences $A=(A_0,A_{1}(1),\ldots,A_{n}(1,\ldots,n),\ldots)$ of self-adjoint operators
$A_{n}\in\mathfrak{L}(\mathcal{H}_{n})$.

In this case the mean value (expectation value) of an observable is determined by the
positive continuous linear functional, which is represented by the following series 
expansion:
\begin{eqnarray}\label{averageD}
     &&\langle A\rangle=(I,D)^{-1}\sum\limits_{n=0}^{\infty}\frac{1}{n!}
         \,\mathrm{Tr}_{1,\ldots,n}\,A_{n}\,D_{n},
\end{eqnarray}
where the sequence $D=(I,D_{1},\ldots,D_{n},\ldots)$ of self-adjoint positive operators
$D_{n}\in\mathfrak{L}^1(\mathcal{H}_{n})$ is a sequence of density operators, describing
all possible states of a quantum system of non-fixed number of particles, and
$(I,D)={\sum\limits}_{n=0}^{\infty}\frac{1}{n!}\mathrm{Tr}_{1,\ldots,n}D_{n}$ 
is a normalization factor. The functional \eqref{averageD} that defines a duality of
observables and states, exists if if $D_{n}\in\mathfrak{L}^{1}(\mathcal{H}_{n})$ and
$A_{n}\in\mathfrak{L}(\mathcal{H}_{n})$.

We note that in case of a system of fixed number $N<\infty$ of particles the observables
and states are one-component sequences $A^{(N)}=(0,\ldots,0,A_{N},0,\ldots)$ and
$D^{(N)}=(0,\ldots,0,D_{N},0,\ldots)$, respectively, and hence, mean value functional
\eqref{averageD} takes the conventional representation
\begin{eqnarray*}
   &&\langle A^{(N)}\rangle=(\mathrm{Tr}_{1,\ldots,N}D_{N})^{-1}\mathrm{Tr}_{1,\ldots,N}A_{N}D_{N},
\end{eqnarray*}
and it is usually assumed that the normalization condition $\mathrm{Tr}_{1,\ldots,N}D_{N}=1$, holds.

If initial state is specified by the sequence of density operators
$D(0)=(I,D_{1}^{0}(1),\ldots,D_{n}^{0}(1,\ldots,n),\ldots)$, then the evolution of all possible
states, i.e. the sequence $D(t)=(I,D_{1}(t,1),\ldots,D_{n}(t,1,\ldots,n),\ldots)$ of the density
operators $D_{n}(t),\,n\geq1$, is determined by the following groups of operators:
\begin{eqnarray}\label{rozv_fon-N}
    &&D_n(t)=\mathcal{G}^{\ast}_n(t)D_n^{0}\doteq e^{-itH_{n}}D_n^{0}e^{itH_{n}}, \quad n\geq1,
\end{eqnarray}
where the self-adjoint operator $H_{n}$ is the $n$-particle Hamiltonian, and we used units where 
$h={2\pi\hbar}=1$ is a Planck constant. In the sense of the mean value functional \eqref{averageD},
the group $\mathcal{G}_{n}^{\ast}(t)$ is conjugated to the group of operators $\mathcal{G}_{n}(t)$,
which describes the evolution of the observables.

The one-parameter mapping $\mathcal{G}^{\ast}_n(t)$ is defined on the space of trace class
operators $\mathfrak{L}^1(\mathcal{H}_{n})$, and it is an isometric strongly continuous
group of operators that preserves positivity and self-adjointness of operators \cite{DauL}.
In the sequel the inverse group to the group $\mathcal{G}_{n}^{\ast}(t)$ we will denote by
symbol $(\mathcal{G}_{n}^{\ast})^{-1}(t)=\mathcal{G}_{n}^{\ast}(-t)$. On its domain of the
definition the infinitesimal generator $\mathcal{N}^{\ast}_{n}$ of the group of operators
$\mathcal{G}^{\ast}_n(t)$ is determined in the sense of the strong convergence of the space
$\mathfrak{L}^1(\mathcal{H}_{n})$ by the generator of the von Neumann equation (quantum
Liouville equation), namely
\begin{eqnarray}\label{infOper1}
    &&\lim\limits_{t\rightarrow 0}\frac{1}{t}\big(\mathcal{G}^{\ast}_n(t)f_n-f_n \big)
    =-i\,(H_n f_n - f_n H_n)\doteq\mathcal{N}^{\ast}_n f_n.
\end{eqnarray}
The operator $\mathcal{N}^{\ast}_n$  has the following structure:
$\mathcal{N}^{\ast}_n=\sum_{j=1}^{n}\mathcal{N}^{\ast}(j)+
\sum_{j_{1}<j_{2}=1}^{n}\mathcal{N}^{\ast}_{\mathrm{int}}(j_{1},j_{2})$,
where the operator $\mathcal{N}^{\ast}(j)$ is a free motion generator of the von Neumann
equation, the operator $\mathcal{N}^{\ast}_{\mathrm{int}}$ is defined by means of the
operator of a two-body interaction potential $\Phi$ by the formula:
$\mathcal{N}^{\ast}_{\mathrm{int}}(j_{1},j_{2})f_n\doteq -i\,(\Phi(j_{1},j_{2})f_n-f_n \Phi(j_{1},j_{2}))$.

If $D_{n}^{0}\in\mathfrak{L}^{1}(\mathcal{H}_{n}),\, n\geq1$, then for $t\in\mathbb{R}$
the sequence of density operators \eqref{rozv_fon-N} is a unique solution of the Cauchy
problem of the von Neumann equations \cite{DauL}:
\begin{eqnarray}
  \label{vonNeumannEqn}
     &&\frac{\partial}{\partial t}D_{n}(t)=\mathcal{N}^{\ast}_{n}\,D_{n}(t),\\
  \label{F-N12}
     &&D(t)_{n}\big|_{t=0}=D_n^{0}, \quad n\geq1,
\end{eqnarray}
where the generators $\mathcal{N}^{\ast}_{n},\, n\geq1,$ of equations \eqref{vonNeumannEqn}
are adjoint operators to generators of the Heisenberg equations for observables in the
sense of mean value functional \eqref{averageD}, and are defined by formula \eqref{infOper1}.

We remind that the density operator is represented as a convex linear combination of a rank
one projectors, though this representation is not unique. A density operator that is a rank
one projection $D_{n}(t)=P_{\psi_n}(t), \,\psi_n\in\mathcal{H}_{n},$ is known as a pure quantum
state, and all quantum states that are not pure are interpreted as mixed states.

In consequence of the validity for the projector of the equality $P_{\psi_n}(t)=P_{\psi_n(t)}$,
where $\psi_n(t)=e^{-itH_{n}}\psi_n$, we conclude that the evolution of pure states can be also
described by means of the Cauchy problem for the Schr\"{o}dinger equation:
\begin{eqnarray*}
  \label{Sch}
     &&i\frac{\partial}{\partial t}\psi_{n}(t)=H_{n}\psi_{n}(t),\\
  \label{Schi}
     &&\psi(t)_{n}\big|_{t=0}=\psi_n, \quad n\geq1,
\end{eqnarray*}
where the operator $H_{n}$ is the $n$-particle Hamiltonian.

\section{Cluster expansions of the density operator}

An alternative approach to the description of states of a quantum system of finitely
many particles is given by means of operators determined by the cluster expansions
of the density operator. They are interpreted as correlation operators.

We introduce the sequence of correlation operators $g(t)=(I,g_{1}(t,1),\ldots,g_{s}(t,$
$1,\ldots,s),\ldots)$ using cluster expansions of a sequence of density operators
$D(t)=(I,D_{1}(t,1),\ldots,D_{n}(t,1,\ldots,n),\ldots)$:
\begin{eqnarray}\label{D_(g)N}
    &&D_{n}(t,1,\ldots,n)= g_{n}(t,1,\ldots,n)+\nonumber\\
    &&\sum\limits_{\mbox{\scriptsize $\begin{array}{c}\mathrm{P}:
    (1,\ldots,n)=\bigcup_{i}X_{i},\\|\mathrm{P}|>1 \end{array}$}}
        \prod_{X_i\subset \mathrm{P}}g_{|X_i|}(t,X_i),\quad n\geq1,
\end{eqnarray}
where ${\sum\limits}_{\mathrm{P}:(1,\ldots,n)=\bigcup_{i} X_{i},\,|\mathrm{P}|>1}$ is the sum
over all possible partitions $\mathrm{P}$ of the set $(1,\ldots,n)$ into $|\mathrm{P}|>1$
nonempty mutually disjoint subsets $X_i\subset (1,\ldots,n)$.

Solutions of recursion relations \eqref{D_(g)N} are given by the following expansions:
\begin{eqnarray}\label{gfromDFB}
   &&g_{s}(t,1,\ldots,s)=D_{s}(t,1,\ldots,s)+\nonumber\\
   &&\sum\limits_{\mbox{\scriptsize $\begin{array}{c}\mathrm{P}:(1,\ldots,s)=
       \bigcup_{i}X_{i},\\|\mathrm{P}|>1\end{array}$}}(-1)^{|\mathrm{P}|-1}(|\mathrm{P}|-1)!\,
       \prod_{X_i\subset \mathrm{P}}D_{|X_i|}(t,X_i), \quad s\geq1.
\end{eqnarray}
The structure of expansions \eqref{gfromDFB} is such that the correlation operators can be treated
as cumulants (semi-invariants) of the density operators \eqref{rozv_fon-N}.

Thus, correlation operators \eqref{gfromDFB} are to enable to describe of the evolution of states
of finitely many particles by the equivalent method in comparison with the density operators, namely
within the framework of dynamics of correlations \cite{GerS},\cite{GP11}.

If initial state described by the sequence of correlation operators
$g(0)=(I,g_{1}^{0}(1),\ldots,$ $g_{n}^{0}(1,\ldots,n),\ldots)\in\oplus_{n=0}^{\infty}\mathfrak{L}^{1}(\mathcal{H}_{n})$,
then the evolution of all possible states, i.e. the sequence $g(t)=(I,g_{1}(t,1),\ldots,g_{s}(t,1,\ldots,s),\ldots)$
of the correlation operators $g_{s}(t),\,s\geq1$, is determined by the following group of nonlinear operators \cite{GP11}:
\begin{eqnarray}\label{ghs}
   &&g(t,1,\ldots,s)=\mathcal{G}(t;1,\ldots,s\mid g(0))\doteq\\
   &&\sum\limits_{\mathrm{P}:\,(1,\ldots,s)=\bigcup_j X_j}
      \mathfrak{A}_{|\mathrm{P}|}(t,\{X_1\},\ldots,\{X_{|\mathrm{P}|}\})
      \prod_{X_j\subset \mathrm{P}}g_{|X_j|}^{0}(X_j),\quad s\geq1,\nonumber
\end{eqnarray}
where $\sum_{\mathrm{P}:\,(1,\ldots,s)=\bigcup_j X_j}$ is the sum over all
possible partitions $\mathrm{P}$ of the set $(1,\ldots,s)$ into $|\mathrm{P}|$ nonempty
mutually disjoint subsets $X_j$, the set $(\{X_1\},\ldots,\{X_{|\mathrm{P}|}\})$ consists
from elements of which are subsets $X_j\subset (1,\ldots,s)$, i.e.
$|(\{X_1\},\ldots,\{X_{|\mathrm{P}|}\})|=|\mathrm{P}|$.
The generating operator $\mathfrak{A}_{|\mathrm{P}|}(t)$ in expansion \eqref{ghs} is the
$|\mathrm{P}|th$-order cumulant of the groups of operators \eqref{rozv_fon-N} which is
defined by the expansion
\begin{eqnarray}\label{cumulantP}
   &&\hskip-8mm \mathfrak{A}_{|\mathrm{P}|}(t,\{X_1\},\ldots,\{X_{|\mathrm{P}|}\})\doteq\\
   &&\hskip-8mm \sum\limits_{\mathrm{P}^{'}:\,(\{X_1\},\ldots,\{X_{|\mathrm{P}|}\})=
      \bigcup_k Z_k}(-1)^{|\mathrm{P}^{'}|-1}({|\mathrm{P}^{'}|-1})!
      \prod\limits_{Z_k\subset\mathrm{P}^{'}}\mathcal{G}^{\ast}_{|\theta(Z_{k})|}(t,\theta(Z_{k})),\nonumber
\end{eqnarray}
where $\theta$ is the declusterization mapping: $\theta(\{X_1\},\ldots,\{X_{|\mathrm{P}|}\})\doteq(1,\ldots,s)$.

In particular case of the absence of correlations between particles at the initial time (known as
initial states satisfying a chaos condition \cite{BPSh},\cite{Sh},\cite{BCEP07}) the
sequence of initial correlation operators has the form $g^{c}(0)=(0,g_{1}^{0}(1),0,\ldots,0,\ldots)$
(in case of the Maxwell--Boltzmann statistics in terms of a sequence of density operators it means
that $D^{c}(0)=(I,D_{1}^0(1),D_{1}^0(1)D_{1}^0(2),\ldots,\prod^n_{i=1} D_{1}^0(i),\ldots)$).
In this case expansions \eqref{ghs} are represented as follows:
\begin{eqnarray*}\label{gth}
   &&g_{s}(t,1,\ldots,s)=\mathfrak{A}_{s}(t,1,\ldots,s)\,\prod\limits_{i=1}^{s}g_{1}^{0}(i),\quad s\geq1,
\end{eqnarray*}
where $\mathfrak{A}_{s}(t)$ is the $sth$-order cumulant of groups of operators
\eqref{rozv_fon-N} defined by the expansion
\begin{eqnarray}\label{cumcp}
   &&\mathfrak{A}_{s}(t,1,\ldots,s)=\sum\limits_{\mathrm{P}:\,(1,\ldots,s)=
       \bigcup_i X_i}(-1)^{|\mathrm{P}|-1}({|\mathrm{P}|-1})!
      \prod\limits_{X_i\subset\mathrm{P}}\mathcal{G}^{\ast}_{|X_i|}(t,X_i),
\end{eqnarray}
and it was used notations accepted in formula \eqref{rozv_fon-N}.

If $g_{s}^{0}\in\mathfrak{L}^{1}(\mathcal{H}_{s}),\, s\geq1$, then for $t\in\mathbb{R}$ the sequence
of correlation operators \eqref{ghs} is a unique solution of the Cauchy problem of the quantum
von Neumann hierarchy \cite{GerS},\cite{GP11}:
\begin{eqnarray}\label{vNh}
   &&\hskip-5mm\frac{\partial}{\partial t}g_{s}(t,1,\ldots,s)=\mathcal{N}^{\ast}_{s}g_{s}(t,1,\ldots,s)+\\
   &&\hskip-5mm\sum\limits_{\mathrm{P}:\,(1,\ldots,s)=X_{1}\bigcup X_2}\,\sum\limits_{i_{1}\in X_{1}}
      \sum\limits_{i_{2}\in X_{2}}\mathcal{N}_{\mathrm{int}}^{\ast}(i_{1},i_{2})
      g_{|X_{1}|}(t,X_{1})g_{|X_{2}|}(t,X_{2}), \nonumber\\
   \nonumber\\
 \label{vNhi}
   &&\hskip-5mm g_{s}(t,1,\ldots,s)\big|_{t=0}=g_{s}^{0}(1,\ldots,s),\quad s\geq1,
\end{eqnarray}
where ${\sum\limits}_{\mathrm{P}:\,(1,\ldots,s)=X_{1}\bigcup X_2}$ is the sum over
all possible partitions $\mathrm{P}$ of the set $(1,\ldots,s)$ into two nonempty mutually disjoint
subsets $X_1$ and $X_2$, and the operator $\mathcal{N}^{\ast}_{s}$ is defined on the subspace
$\mathfrak{L}^{1}_0(\mathcal{H}_s)$ by formula \eqref{infOper1}. It should be noted that the
von Neumann hierarchy \eqref{vNh} is the evolution recurrence equations set.

\section{Reduced density operators}

For the description of quantum systems of both finite and infinite number of particles
another approach to describe of states and observables is used, which is equivalent to
the approach formulated above in case of systems of finitely many particles \cite{BogLect},\cite{CGP97}.

Indeed, for a system of finitely many particles mean value functional \eqref{averageD} can be
represented in one more form
\begin{eqnarray}\label{avmar}
       &&\langle A\rangle=(I,D)^{-1}\sum\limits_{n=0}^{\infty}\frac{1}{n!}\,
         \mathrm{Tr}_{1,\ldots,n}\,A_{n}\,D_{n}=\\
       &&\sum\limits_{s=0}^{\infty}\frac{1}{s!}\,
         \mathrm{Tr}_{1,\ldots,s}\,B_{s}(1,\ldots,s)\,F_{s}(1,\ldots,s),\nonumber
\end{eqnarray}
where, for the description of observables and states, the sequence of so-called reduced observables
$B=(B_0,B_{1}(1),\ldots,B_{s}(1,\ldots,s),\ldots)$
(other used terms: marginal or $s$-particle observable) was introduced and reduced density operators
$F=(I,F_{1}(1),\ldots,F_{s}(1,\ldots,s),\ldots)$ (other used terms: marginal or $s$-particle density
operators \cite{BogLect},\cite{BPSh}), respectively. Thus, the reduced observables are defined by means
of observables by the following expansions \cite{BG},\cite{G11}:
\begin{eqnarray}\label{mo}
    &&\hskip-5mm B_{s}(1,\ldots,s)\doteq\sum_{n=0}^s\,\frac{(-1)^n}{n!}\sum_{j_1\neq\ldots\neq j_{n}=1}^s
          A_{s-n}((1,\ldots,s)\setminus(j_1,\ldots,j_{n})), \,\, s\geq 1,
\end{eqnarray}
and the reduced density operators are defined by means of density operators as follows \cite{CGP97}
\begin{eqnarray}\label{ms}
     &&\hskip-5mm F_{s}(1,\ldots,s)\doteq(I,D)^{-1}\sum\limits_{n=0}^{\infty}\frac{1}{n!}
         \mathrm{Tr}_{s+1,\ldots,s+n}\,D_{s+n}(1,\ldots,s+n), \,\, s\geq 1.
\end{eqnarray}

We emphasize that the possibility of describing states within the framework of reduced density
operators naturally arises as a result of dividing the series in expression \eqref{averageD}
by the series of the normalization factor, i.e. in consequence of redefining of mean value
functional \eqref{avmar}.

If initial state specified by the sequence of reduced density operators
$F(0)=(I,F_{1}^{0}(1),\ldots,F_{n}^{0}(1,\ldots,n),\ldots)$, then the evolution of all possible
states, i.e. a sequence $F(t)=(I,F_{1}(t,1),\ldots,F_{s}(t,1,\ldots,s),\ldots)$ of the reduced
density operators $F_{s}(t),\,s\geq1$, is determined by the following series expansion \cite{GerS06},\cite{GerRS}:
\begin{eqnarray}\label{RozvBBGKY}
   &&\hskip-5mm F_{s}(t,1,\ldots,s)=\\
   &&\hskip-5mm \sum\limits_{n=0}^{\infty}\frac{1}{n!}\,\mathrm{Tr}_{s+1,\ldots,{s+n}}\,
       \mathfrak{A}_{1+n}(t,\{1,\ldots,s\},s+1,\ldots,{s+n})F_{s+n}^{0}(1,\ldots,{s+n}),\nonumber\\
   &&\hskip-5mm  s\geq1,\nonumber
\end{eqnarray}
where the generating operator
\begin{eqnarray}\label{cumulant1+n}
   &&\mathfrak{A}_{1+n}(t,\{1,\ldots,s\},s+1,\ldots,{s+n})=\\
   &&\sum\limits_{\mathrm{P}\,:(\{1,\ldots,s\},s+1,\ldots,{s+n})=
      {\bigcup\limits}_i X_i}(-1)^{|\mathrm{P}|-1}(|\mathrm{P}|-1)!
      \prod_{X_i\subset\mathrm{P}}\mathcal{G}^{\ast}_{|\theta(X_i)|}(t,\theta(X_i))\nonumber
\end{eqnarray}
is the $(1+n)th$-order cumulant of groups of operators \eqref{rozv_fon-N} \cite{GerRS}.
In expansion \eqref{cumulant1+n} the symbol ${\sum\limits}_\mathrm{P}$ means the sum over
all possible partitions $\mathrm{P}$ of the set $(\{1,\ldots,s\},s+1,\ldots,{s+n})$ into
$|\mathrm{P}|$ nonempty mutually disjoint subsets $X_i\subset(\{1,\ldots,s\},s+1,\ldots,{s+n})$
and we use notations accepted in formula \eqref{ghs}.

If $F(0)\in\oplus_{n=0}^{\infty}\alpha^{n}\mathfrak{L}^{1}(\mathcal{H}_{n})$ and $\alpha>e$,
then for $t\in\mathbb{R}$ the sequence of reduced density operators \eqref{RozvBBGKY} is
a unique solution of the Cauchy problem of the quantum BBGKY hierarchy \cite{BogLect}:
\begin{eqnarray}
 \label{BBGKY}
   &&\frac{\partial}{\partial t}F_{s}(t,1,\ldots,s)=\mathcal{N}^{\ast}_{s}F_{s}(t,1,\ldots,s)+ \\
   &&\sum\limits_{j=1}^{s}\mathrm{Tr}_{s+1}\mathcal{N}^{\ast}_{\mathrm{int}}(j,s+1)F_{s+1}(t,1,\ldots,s,s+1),\nonumber\\
       \nonumber \\
 \label{BBGKYi}
   &&F_{s}(t,1,\ldots,s)\mid_{t=0}=F_{s}^{0}(1,\ldots,s),\quad s\geq 1,
\end{eqnarray}
where we used notations accepted in formula \eqref{infOper1}.

We note that traditionally \cite{BogLect},\cite{BPSh},\cite{CGP97},\cite{Go16} the reduced density
operators are represented by means of the perturbation theory series of the BBGKY hierarchy \eqref{BBGKY}
\begin{eqnarray*}\label{iter}
   &&\hskip-7mm F_s(t,1,\ldots,s)=\\
   &&\hskip-7mm \sum\limits_{n=0}^{\infty}\,\int\limits_{0}^{t}dt_{1}\ldots
       \int\limits_{0}^{t_{n-1}}dt_{n}\mathrm{Tr}_{s+1,\ldots,s+n}\mathcal{G}^{\ast}_s(t-t_{1})
       \sum\limits_{j_1=1}^{s}\mathcal{N}^{\ast}_{\mathrm{int}}(j_1,s+1))
       \mathcal{G}^{\ast}_{s+1}(t_1-t_2)\ldots\nonumber\\
   &&\hskip-7mm\mathcal{G}^{\ast}_{s+n-1}(t_{n-1}-t_n)
       \sum\limits_{j_n=1}^{s+n-1}\mathcal{N}^{\ast}_{\mathrm{int}}(j_n,s+n))
       \mathcal{G}^{\ast}_{s+n}(t_{n})F_{s+n}^0(1,\ldots,s+n), \, s\geq1,\nonumber
\end{eqnarray*}
where we used notations accepted in formula \eqref{infOper1}. The nonperturbative series expansion
for reduced density operators \eqref{RozvBBGKY} is represented in the form of the perturbation
theory series for suitable interaction potentials and initial data as a result of the employment of analogs
of the Duhamel equation to cumulants \eqref{cumulant1+n} of the groups of operators \eqref{rozv_fon-N}.

An equivalent definition of reduced density operators can be formulated based on correlation
operators \eqref{ghs} of systems of finitely many particles \cite{GP11}, namely
\begin{eqnarray}\label{FClusters}
    &&\hskip-12mm F_{s}(t,1,\ldots,s)\doteq\sum\limits_{n=0}^{\infty}\frac{1}{n!}\,
       \mathrm{Tr}_{s+1,\ldots,s+n}\,\,g_{1+n}(t,\{1,\ldots,s\},s+1,\ldots,s+n), \, s\geq1,
\end{eqnarray}
where the correlation operators of clusters of particles $g_{1+n}(t), n\geq0,$ are defined
by the expansions
\begin{eqnarray}\label{rozvNF-Nclusters}
    &&\hskip-7mm g_{1+n}(t,\{1,\ldots,s\},s+1,\ldots,s+n)=\\
    &&\hskip-7mm \sum\limits_{\mathrm{P}:\,(\{1,\ldots,s\},\,s+1,\ldots,s+n)=\bigcup_i X_i}
       \mathfrak{A}_{|\mathrm{P}|}\big(-t,\{\theta(X_1)\},\ldots,\{\theta(X_{|\mathrm{P}|})\}\big)
       \prod_{X_i\subset \mathrm{P}}g_{|X_i|}^0(X_i),\nonumber\\
     &&\hskip-7mm n\geq0,\nonumber
\end{eqnarray}
and $\mathfrak{A}_{|\mathrm{P}|}(t)$ is the $|\mathrm{P}|th$-order cumulant \eqref{cumulantP}
of the groups of operators \eqref{rozv_fon-N}. Owing that correlation operators $g_{1+n}(t),\,n\geq0,$
are governed by the corresponding von Neumann hierarchy, for reduced density operators \eqref{FClusters}
we can derive the quantum BBGKY hierarchy.

Thus, as follows from the above, the cumulant structure of correlation operator expansion
\eqref{rozvNF-Nclusters} induces the cumulant structure of series expansions for reduced
density operators \eqref{RozvBBGKY}, i.e. in fact, dynamics of correlations is generated
dynamics of infinitely many particles.

\section{Reduced correlation operators}

Another approach to the description of states of quantum systems of both finite and
infinite number of particles is can be formulated as in above  by means of operators
determined by the cluster expansions of the reduced density operators. Such operators
are interpreted as reduced correlation operators of states (marginal or $s$-particle
correlation operators) \cite{BogLect},\cite{GP13},\cite{G17}.

Traditionally reduced correlation operators are introduced by means of the cluster expansions
of the reduced density operators \eqref{FClusters} as follows:
\begin{eqnarray}\label{FG}
   &&F_{s}(t,1,\ldots,s)=
      \sum\limits_{\mbox{\scriptsize$\begin{array}{c}\mathrm{P}:(1,\ldots,s)=\bigcup_{i}X_{i}\end{array}$}}
      \prod_{X_i\subset \mathrm{P}}G_{|X_i|}(t,X_i), \quad s\geq1,
\end{eqnarray}
where ${\sum\limits}_{\mathrm{P}:(1,\ldots,s)=\bigcup_{i} X_{i}}$ is the sum over all possible
partitions $\mathrm{P}$ of the set $(1,\ldots,s)$ into $|\mathrm{P}|$ nonempty mutually disjoint
subsets $X_i\subset(1,\ldots,s)$. As a consequence of this, the solution of recurrence relations
\eqref{FG} represented through reduced density operators as follows
\begin{eqnarray}\label{gBigfromDFB}
   &&\hskip-7mm G_{s}(t,1,\ldots,s)=
     \sum\limits_{\mbox{\scriptsize $\begin{array}{c}\mathrm{P}:(1,\ldots,s)=\bigcup_{i}X_{i}\end{array}$}}
     (-1)^{|\mathrm{P}|-1}(|\mathrm{P}|-1)!\,\prod_{X_i\subset \mathrm{P}}F_{|X_i|}(t,X_i), \\
   &&\hskip-7mm   s\geq1,\nonumber
\end{eqnarray}
are interpreted as the operators that describe correlations of states in many-particle systems.
The structure of expansions \eqref{gBigfromDFB} is such that the reduced correlation operators
can be treated as cumulants (semi-invariants) of the reduced density operators \eqref{RozvBBGKY}.

Assuming as a basis an alternative approach to the description of the evolution of states of
quantum many-particle systems within the framework of correlation operators \eqref{ghs},
we can define the reduced correlation operators by means of a solution of the Cauchy problem
of the von Neumann hierarchy \eqref{vNh},\eqref{vNhi} as follows \cite{GP13},\cite{G17}:
\begin{eqnarray}\label{Gexpg}
   &&G_{s}(t,1,\ldots,s)\doteq\sum\limits_{n=0}^{\infty}\frac{1}{n!}\,
      \mathrm{Tr}_{s+1,\ldots,s+n}\,\,g_{s+n}(t,1,\ldots,s+n), \quad s\geq1,
\end{eqnarray}
where the operator $g_{s+n}(t,1,\ldots,s+n)$ is defined by expansion \eqref{gfromDFB}.
We emphasize that every term of the expansion \eqref{Gexpg} of reduced correlation operator is
determined by the $(s+n)$-particle correlation operator \eqref{ghs} as contrasted to the expansion
of reduced density operator \eqref{FClusters} which is determined by the $(1+n)$-particle correlation
operator of clusters of particles \eqref{rozvNF-Nclusters}.

If $G(0)=(I,G_1^{0}(1),\ldots,G_s^{0}(1,\ldots,s),\ldots)$ is a sequence of reduced correlation
operators at initial instant, then the evolution of all possible states, i.e. a sequence
$G(t)=(I,G_{1}(t,1),\ldots,G_{s}(t,1,\ldots,s),\ldots)$ of the reduced correlation operators
$G_{s}(t),\,s\geq1$, is determined by the following series expansion \cite{G17}:
\begin{eqnarray}\label{sss}
    &&\hskip-12mm G_{s}(t,1,\ldots,s)=\sum\limits_{n=0}^{\infty}\frac{1}{n!}
        \,\mathrm{Tr}_{s+1,\ldots,s+n}\,\mathfrak{A}_{1+n}(t;\{1,\ldots,s\},s+1,\ldots,s+n\mid G(0)), \\
    &&\hskip-12mm s\geq1,\nonumber
\end{eqnarray}
where the generating operator $\mathfrak{A}_{1+n}(t;\{1,\ldots,s\},s+1,\ldots,s+n\mid G(0))$ of this series
is the $(1+n)th$-order cumulant of groups of nonlinear operators \eqref{rozv_fon-N}:
\begin{eqnarray}\label{cc}
   &&\hskip-9mm\mathfrak{A}_{1+n}(t;\{1,\ldots,s\},s+1,\ldots,s+n\mid G(0))\doteq\\
   &&\hskip-9mm\sum\limits_{\mathrm{P}:\,(\{1,\ldots,s\},s+1,\ldots,s+n)=
      \bigcup_k X_k}(-1)^{|\mathrm{P}|-1}({|\mathrm{P}|-1})!
      \mathcal{G}(t;\theta(X_1)\mid\ldots \nonumber\\
    &&\hskip-9mm \mathcal{G}(t;\theta(X_{|\mathrm{P}|})\mid G(0))\ldots), \, n\geq0,\nonumber
\end{eqnarray}
and where the composition of mappings \eqref{rozv_fon-N} of the corresponding noninteracting groups of particles
was denoted by $\mathcal{G}(t;\theta(X_1)\mid \ldots\mathcal{G}(t;\theta(X_{|\mathrm{P}|})\mid G(0))\ldots)$,
for example,
\begin{eqnarray*}
    &&\mathcal{G}\big(t;1\mid\mathcal{G}(t;2\mid G(0))\big)=
        \mathfrak{A}_{1}(t,1)\mathfrak{A}_{1}(t,2)G^{0}_{2}(1,2),\\
    &&\mathcal{G}\big(t;1,2\mid\mathcal{G}(t;3\mid G(0))\big)=
        \mathfrak{A}_{1}(t,\{1,2\})\mathfrak{A}_{1}(t,3)G^{0}_{3}(1,2,3)+\\
    &&\mathfrak{A}_{2}(t,1,2)\mathfrak{A}_{1}(t,3)\big(G^{0}_{1}(1)G^{0}_{2}(2,3)+G^{0}_{1}(2)G^{0}_{2}(1,3)\big).
\end{eqnarray*}

We will adduce examples of expansions \eqref{cc}. The first order cumulant of the groups
of nonlinear operators \eqref{rozv_fon-N} is the group of these nonlinear operators
\begin{eqnarray*}
     &&\mathfrak{A}_{1}(t;\{1,\ldots,s\}\mid G(0))=\mathcal{G}(t;1,\ldots,s\mid G(0)).
\end{eqnarray*}
In case of $s=2$ the second order cumulant of nonlinear operators \eqref{rozv_fon-N}
has the structure
\begin{eqnarray*}
     &&\hskip-8mm \mathfrak{A}_{1+1}(t;\{1,2\},3\mid G(0))=\mathcal{G}(t;1,2,3\mid G(0))-
       \mathcal{G}\big(t;1,2\mid\mathcal{G}(t;3\mid G(0))\big)=\\
     &&\hskip-8mm \mathfrak{A}_{1+1}(t,\{1,2\},3)G^{0}_{3}(1,2,3)+\\
     &&\hskip-8mm \big(\mathfrak{A}_{1+1}(t,\{1,2\},3)-
        \mathfrak{A}_{1+1}(t,2,3)\mathfrak{A}_{1}(t,1)\big)G^{0}_{1}(1)G^{0}_{2}(2,3)+\\
     &&\hskip-8mm\big(\mathfrak{A}_{1+1}(t,\{1,2\},3)-
        \mathfrak{A}_{1+1}(t,1,3)\mathfrak{A}_{1}(t,2)\big)G^{0}_{1}(2)G^{0}_{2}(1,3)+\\
     &&\hskip-8mm \mathfrak{A}_{1+1}(t,\{1,2\},3)G^{0}_{1}(3)G^{0}_{2}(1,2)+
        \mathfrak{A}_{3}(t,1,2,3)G^{0}_{1}(1)G^{0}_{1}(2)G^{0}_{1}(3),
\end{eqnarray*}
where the operator
\begin{equation*}
    \mathfrak{A}_{3}(t,1,2,3)=\mathfrak{A}_{1+1}(t,\{1,2\},3)-\mathfrak{A}_{1+1}(t,2,3)\mathfrak{A}_{1}(t,1)-
     \mathfrak{A}_{1+1}(t,1,3)\mathfrak{A}_{1}(t,2)
\end{equation*}
is cumulant \eqref{cumcp} of groups of operators \eqref{rozv_fon-N} of the third order.

In the case of the initial state specified by the sequence of reduced correlation operators
$G^{(c)}=(0,G_1^{0},0,\ldots,0,\ldots)$, that is, in the absence of correlations between particles
at the initial moment of time \cite{BPSh},\cite{Sh},\cite{BCEP07}, according to definition \eqref{cc},
reduced correlation operators \eqref{sss} are represented by the following series expansions:
\begin{eqnarray}\label{mcc}
   &&\hskip-5mm G_{s}(t,1,\ldots,s)=\sum\limits_{n=0}^{\infty}\frac{1}{n!}
       \,\mathrm{Tr}_{s+1,\ldots,s+n}\,\mathfrak{A}_{s+n}(t;1,\ldots,s+n)
       \prod_{i=1}^{s+n}G_1^{0}(i), \, s\geq1,
\end{eqnarray}
where the generating operator $\mathfrak{A}_{s+n}(t)$ is the $(s+n)th$-order cumulant \eqref{cumcp}
of groups of operators \eqref{rozv_fon-N}.

If $G(0)\in\oplus_{n=0}^{\infty}\mathfrak{L}^{1}(\mathcal{H}_{n})$, then for $t\in\mathbb{R}$
the sequence of reduced correlation operators \eqref{sss} is a unique solution of the Cauchy problem
of the hierarchy of nonlinear evolution equations (known as the nonlinear quantum BBGKY hierarchy) \cite{G17}:
\begin{eqnarray}\label{gBigfromDFBa}
   &&\hskip-8mm\frac{\partial}{\partial t}G_s(t,1,\ldots,s)=\mathcal{N}^{\ast}_{s}G_{s}(t,1,\ldots,s)+\\
   &&\hskip-8mm \sum\limits_{\mathrm{P}:\,(1,\ldots,s)=X_{1}\bigcup X_2}\,\sum\limits_{i_{1}\in X_{1}}
      \sum\limits_{i_{2}\in X_{2}}\mathcal{N}_{\mathrm{int}}^{\ast}(i_{1},i_{2})
      G_{|X_{1}|}(t,X_{1})G_{|X_{2}|}(t,X_{2}))+\nonumber\\
   &&\hskip-8mm\mathrm{Tr}_{s+1}
      \sum_{i\in Y}\mathcal{N}^{\ast}_{\mathrm{int}}(i,s+1)\big(G_{s+1}(t,1,\ldots,s+1)+\nonumber\\
   &&\hskip-8mm \sum_{\mbox{\scriptsize$\begin{array}{c}\mathrm{P}:(1,\ldots,s+1)=X_1\bigcup X_2,\\i\in
      X_1;s+1\in X_2\end{array}$}}G_{|X_1|}(t,X_1)G_{|X_2|}(t,X_2)\big),\nonumber\\ \nonumber\\
 \label{gBigfromDFBai}
   &&\hskip-8mmG_{s}(t,1,\ldots,s)\big|_{t=0}=G_{s}^{0}(,1,\ldots,s), \, s\geq1,
\end{eqnarray}
where we use accepted in hierarchy \eqref{vNh} notations.

We note that the reduced correlation operators give an equivalent approach to the
description of the evolution of states of quantum many-particle systems as compared with
the reduced density operators. Indeed, the macroscopic characteristics of fluctuations
of observables are directly determined by the reduced correlation operators on the microscopic
scale \cite{BogLect},\cite{GP13}, for example, the functional of the dispersion of an
additive-type observable, i.e. the sequence  $A^{(1)}=(0,a_{1}(1),\ldots,\sum_{i_{1}=1}^{n}a_1(i_{1}),\ldots)$,
is represented by the formula
\begin{eqnarray*}
    &&\hskip-8mm \langle(A^{(1)}-\langle A^{(1)}\rangle)^2\rangle(t)=
      \mathrm{Tr}_{1}\,(a_1^2(1)-\langle A^{(1)}\rangle^2(t))G_{1}(t,1)+
      \mathrm{Tr}_{1,2}\,a_{1}(1)a_{1}(2)G_{2}(t,1,2),
\end{eqnarray*}
where $\langle A^{(1)}\rangle(t)=\mathrm{Tr}_{1}\,a_{1}(1)G_{1}(t,1)$ is the mean value functional
of an additive-type observable.

\section{On the description of the evolution of states by the one-particle correlation operator}

Further, we shall consider systems which the initial state specified by a one-particle reduced
correlation (density) operator, namely, the initial state specified by a sequence of reduced
correlation operators satisfying a chaos property stated above, i.e. by the sequence
$G^{(c)}=(0,G_1^{0},0,\ldots,0,\ldots)$. We remark that such an assumption about initial states
is intrinsic in kinetic theory of many-particle systems.

The following statement is true. In the case of the initial state specified by a one-particle
correlation (density) operator $G^{(c)}$ the evolution that described within the framework of
the sequence $G(t)=\left(I,G_1(t),\ldots,G_s(t),\ldots\right)$ of reduced correlation operators
\eqref{sss}, is also be described by the sequence
$G(t\mid G_{1}(t))=(I,G_1(t),G_2(t\mid G_{1}(t)),\ldots,G_s(t\mid G_{1}(t)),\ldots)$ of reduced 
(marginal) correlation functionals: $G_s(t,1,\ldots,s\mid G_{1}(t)),\,s\geq2$, with respect to 
the one-particle correlation operator $G_1(t)$ governed by the generalized quantum kinetic 
equation \cite{G17IM},\cite{GT}.

In the case under consideration the reduced correlation functionals
$G_s(t\mid G_{1}(t)),\,s\geq2$, are represented with respect to the one-particle
correlation operator
\begin{eqnarray}\label{ske}
   &&\hskip-8mm G_{1}(t,1)=\sum\limits_{n=0}^{\infty}\frac{1}{n!}\,\mathrm{Tr}_{2,\ldots,{1+n}}\,
      \mathfrak{A}_{1+n}(t, 1,\ldots, n+1)\prod_{i=1}^{n+1}G_{1}^{0}(i),
\end{eqnarray}
where the generating operator $\mathfrak{A}_{1+n}(t)$ is cumulant \eqref{cumcp}
of the groups of operators \eqref{rozv_fon-N} of the $(1+n)th$-order, by the following series:
\begin{eqnarray}\label{f}
     &&\hskip-12mm G_{s}\bigl(t,1,\ldots,s\mid G_{1}(t)\bigr)=\\
     &&\hskip-12mm \sum _{n=0}^{\infty }\frac{1}{n!}\,\mathrm{Tr}_{s+1,\ldots,{s+n}}\,
        \mathfrak{V}_{s+n}\bigl(t,\theta(\{1,\ldots,s\}),s+1,\ldots,s+n\bigr)\prod_{i=1}^{s+n}G_{1}(t,i),
        \,\, s\geq2.\nonumber
\end{eqnarray}
The generating operator $\mathfrak{V}_{s+n}(t),\,n\geq0$, of the $(s+n)th$-order of this series
is determined by the following expansion \cite{GT}
\begin{eqnarray}\label{skrrc}
   &&\hskip-7mm\mathfrak{V}_{s+n}\bigl(t,\theta(\{1,\ldots,s\}),s+1,\ldots,s+n\bigr)=\\
   &&\hskip-7mm n!\,\sum_{k=0}^{n}\,(-1)^k\,\sum_{n_1=1}^{n}\ldots
       \sum_{n_k=1}^{n-n_1-\ldots-n_{k-1}}\frac{1}{(n-n_1-\ldots-n_k)!}\times\nonumber\\
   &&\hskip-7mm \hat{\mathfrak{A}}_{s+n-n_1-\ldots-n_k}(t,\theta(\{1,\ldots,s\}),s+1,\ldots,
       s+n-n_1-\ldots-n_k)\times\nonumber\\
   &&\hskip-7mm \prod_{j=1}^k\,\sum\limits_{\mbox{\scriptsize$\begin{array}{c}
       \mathrm{D}_{j}:Z_j=\bigcup_{l_j}X_{l_j},\\
       |\mathrm{D}_{j}|\leq s+n-n_1-\dots-n_j\end{array}$}}\frac{1}{|\mathrm{D}_{j}|!}
       \sum_{i_1\neq\ldots\neq i_{|\mathrm{D}_{j}|}=1}^{s+n-n_1-\ldots-n_j}\,
       \prod_{X_{l_j}\subset \mathrm{D}_{j}}\,\frac{1}{|X_{l_j}|!}\hat{\mathfrak{A}}_{1+|X_{l_j}|}(t,i_{l_j},X_{l_j}).\nonumber
\end{eqnarray}
where $\sum_{\mathrm{D}_{j}:Z_j=\bigcup_{l_j} X_{l_j}}$ is the sum over all possible dissections
\cite{GT} of the linearly ordered set $Z_j\equiv(s+n-n_1-\ldots-n_j+1,\ldots,s+n-n_1-\ldots-n_{j-1})$
on no more than $s+n-n_1-\ldots-n_j$ linearly ordered subsets, the $(s+n)th$-order scattering cumulant
is defined by the formula
\begin{eqnarray*}
    &&\hskip-8mm\hat{\mathfrak{A}}_{s+n}(t,\theta(\{1,\ldots,s\}),s+1,\ldots,s+n)\doteq
    \mathfrak{A}_{s+n}(t,1,\ldots,s+n)\prod_{i=1}^{s+n}\mathfrak{A}_{1}^{-1}(t,i),
\end{eqnarray*}
and notations accepted above were used. A method of the construction of reduced correlation
functionals \eqref{f} is based on the application of the so-called kinetic cluster expansions \cite{GT}
to the generating operators \eqref{cumcp} of series \eqref{mcc}.

We adduce simplest examples of generating operators \eqref{skrrc}:
\begin{eqnarray*}
   &&\mathfrak{V}_{s}(t,\theta(\{1,\ldots,s\}))=
      \mathfrak{A}_{s}(t,1,\ldots,s)\prod_{i=1}^{s}\mathfrak{A}_{1}^{-1}(t,i),\\
   &&\mathfrak{V}_{s+1}(t,\theta(\{1,\ldots,s\}),s+1)=\mathfrak{A}_{s+1}(t,1,\ldots,s+1)
      \prod_{i=1}^{s+1}\mathfrak{A}_{1}^{-1}(t,i)-\\
   &&\mathfrak{A}_{s}(t,1,\ldots,s)\prod_{i=1}^{s}\mathfrak{A}_{1}^{-1}(t,i)
      \sum_{j=1}^s\mathfrak{A}_{2}(t,j,s+1)\mathfrak{A}_{1}^{-1}(t,j)\mathfrak{A}_{1}^{-1}(t,s+1).
\end{eqnarray*}

We note that reduced correlation functionals \eqref{f} describe all possible correlations
generated by the dynamics of quantum many-particle systems in terms of a one-particle
correlation operator.

If $G_{1}^{0}\in\mathfrak{L}^{1}(\mathcal{H})$, then for arbitrary $t\in\mathbb{R}$ one-particle
correlation operator \eqref{ske} is a weak solution of the Cauchy problem of the generalized
quantum kinetic equation \cite{GT}
\begin{eqnarray}\label{gkec}
   &&\hskip-5mm\frac{\partial}{\partial t}G_{1}(t,1)=\mathcal{N}^{\ast}(1)G_{1}(t,1)+
      \mathrm{Tr}_{2}\,\mathcal{N}_{\mathrm{int}}^{\ast}(1,2)G_{1}(t,1)G_{1}(t,2)+\\
   &&\hskip-5mm \mathrm{Tr}_{2}\,\mathcal{N}_{\mathrm{int}}^{\ast}(1,2)G_{2}\bigl(t,1,2\mid G_{1}(t)\bigr),\nonumber\\
   \nonumber\\
 \label{gkeci}
   &&\hskip-5mm G_{1}(t,1)\big|_{t=0}=G_{1}^{0}(1),
\end{eqnarray}
where the second part of the collision integral in \eqref{gkec} is determined in terms of
the two-particle correlation functional represented by series expansion \eqref{f}.

\section{On the scaling limits of reduced density operators}

The conventional philosophy of the description of the kinetic evolution consists of the following.
If the initial state specified by a one-particle correlation operator, then the evolution of states
can be effectively described by means of a one-particle correlation operator governed by the nonlinear
kinetic equation in a suitable scaling limit.

Further, we consider a scaling asymptotic behavior of the constructed reduced correlation operators
in particular case of a mean field limit for initial states specified by a one-particle
correlation operator mentioned above \cite{BPSh},\cite{Sh},\cite{Go16}.

We will assume the existence of a mean field limit of the initial reduced correlation operator
$G_{1}^{0,\epsilon}$ scaled by the parameter $\epsilon\geq0$ in the following sense
\begin{eqnarray}\label{asic1}
   &&\lim\limits_{\epsilon\rightarrow 0}\big\|\epsilon G_{1}^{0,\epsilon}-
      g_{1}^{0}\big\|_{\mathfrak{L}^{1}(\mathcal{H})}=0,
\end{eqnarray}
and the operator $\mathcal{N}^{\ast}_{\mathrm{int}}$ in hierarchy \eqref{gBigfromDFBa}
scaled in such a way that $\epsilon\mathcal{N}^{\ast}_{\mathrm{int}}$.

Since the $nth$ term of series \eqref{mcc} for the $s$-particle correlation
operator is determined by the $(s+n)th$-order cumulant of asymptotically perturbed groups
of operators \eqref{rozv_fon-N}, then the property of the propagation of initial chaos holds
\begin{eqnarray}\label{Gcid}
   &&\lim\limits_{\epsilon\rightarrow 0}\big\|\epsilon^{s}G_{s}(t)
      \big\|_{\mathfrak{L}^{1}(\mathcal{H}_s)}=0,\quad s\geq2.
\end{eqnarray}

The equality \eqref{Gcid} is derived by the following assertions.
If $f_{s}\in\mathfrak{L}^{1}(\mathcal{H}_{s})$, then for arbitrary finite time interval
for asymptotically perturbed first-order cumulant \eqref{cumcp} of the groups of operators
\eqref{rozv_fon-N}, i.e. for the strongly continuous group \eqref{rozv_fon-N} the following
equality takes place
\begin{eqnarray*}
    &&\lim\limits_{\epsilon\rightarrow 0}\Big\|\mathcal{G}^{\ast}_{s}(t,1,\ldots,s)f_{s}-
        \prod\limits_{j=1}^{s}\mathcal{G}^{\ast}_{1}(t,j)f_{s}\Big\|_{\mathfrak{L}^{1}(\mathcal{H}_{s})}=0.
\end{eqnarray*}
Hence for the $(s+n)th$-order cumulants of asymptotically perturbed groups of operators
\eqref{rozv_fon-N} the following equalities true:
\begin{eqnarray}\label{apg}
   &&\lim\limits_{\epsilon\rightarrow0}\Big\|\frac{1}{\epsilon^{n}}\,
     \mathfrak{A}_{s+n}(t,1,\ldots,s+n)f_{s+n}\Big\|_{\mathfrak{L}^{1}(\mathcal{H}_{s+n})}=0, \,\, s\geq2.
\end{eqnarray}

If for the initial one-particle correlation operator equality (\ref{asic1}) holds, then in case of
$s=1$ for series expansion (\ref{mcc}) the following equality is true
\begin{eqnarray*}
   &&\lim\limits_{\epsilon\rightarrow 0}\big\|\epsilon G_{1}(t)-
     g_{1}(t)\big\|_{\mathfrak{L}^{1}(\mathcal{H})}=0,
\end{eqnarray*}
where for arbitrary finite time interval the limit one-particle correlation operator
$g_1(t,1)$ is represented by the series
\begin{eqnarray}\label{1mco}
   &&\hskip-7mm g_{1}(t,1)=\\
   &&\hskip-7mm\sum\limits_{n=0}^{\infty}\int\limits_0^tdt_{1}\ldots
      \int\limits_0^{t_{n-1}}dt_{n}\,\mathrm{Tr}_{2,\ldots,n+1}\mathcal{G}^{\ast}_{1}(t-t_{1},1)
      \mathcal{N}^{\ast}_{\mathrm{int}}(1,2)\prod\limits_{j_1=1}^{2}
      \mathcal{G}^{\ast}_{1}(t_{1}-t_{2},j_1)\ldots\nonumber\\
   &&\hskip-7mm \prod\limits_{i_{n}=1}^{n}\mathcal{G}^{\ast}_{1}(t_{n}-t_{n},i_{n})
      \sum\limits_{k_{n}=1}^{n}\mathcal{N}^{\ast}_{\mathrm{int}}(k_{n},n+1)\prod\limits_{j_n=1}^{n+1}
      \mathcal{G}^{\ast}_{1}(t_{n},j_n)\prod\limits_{i=1}^{n+1}g_1^{0}(i).\nonumber
\end{eqnarray}

Then we conclude that limit one-particle correlation operator (\ref{1mco})
is a weak solution of the Cauchy problem of the quantum Vlasov kinetic equation
\begin{eqnarray}\label{Vlasov1}
  &&\frac{\partial}{\partial t}g_{1}(t,1)=\mathcal{N}^{\ast}(1)g_{1}(t,1)+
     \mathrm{Tr}_{2}\,\mathcal{N}^{\ast}_{\mathrm{int}}(1,2)g_{1}(t,1)g_{1}(t,2),\\ \nonumber\\
\label{Vlasov1i}
  &&g_{1}(t,1)|_{t=0}=g_{1}^0(1).
\end{eqnarray}

For pure states limit one-particle correlation operator (\ref{1mco}) is governed by the Hartree
equation. Indeed, in terms of the kernel $g_{1}(t,q;q')=\psi(t,q)\psi(t,q')$ of operator (\ref{1mco}),
describing a pure state, in the configuration space representation, kinetic equation (\ref{Vlasov1})
is converted into the Hartree equation
\begin{eqnarray*}
       &&i\frac{\partial}{\partial t}\psi(t,q)=-\frac{1}{2}\Delta_q\psi(t,q)+
          \int dq'\Phi(q-q')|\psi(t,q')|^2\psi(t,q),
\end{eqnarray*}
where the function $\Phi$ is the two-body potential of interaction.

We remark that in case of pure states kinetic equation (\ref{Vlasov1}) can be also transformed
into the nonlinear Schr\"{o}dinger equation \cite{EShY07} or into the Gross--Pitaevskii
kinetic equation \cite{EShY10}.

We remark that some other approaches to the derivation of quantum kinetic equations
\cite{GerUJP}, in particular, quantum systems with initial correlations were developed
in papers \cite{G11},\cite{G15},\cite{G16}.

In the last decade, other scaling limits (weak coupling, low-density, semiclassical)
of the reduced density operators constructed by means theory of perturbations were
rigorously established in numerous papers, for example, in articles \cite{BPSh},\cite{Go16},
\cite{EShY07},\cite{EShY10},\cite{PP09},\cite{GoMP} and papers cited therein.

\section{Conclusion}

This article deals with a quantum system of non-fixed, i.e. arbitrary but finite average
number of identical (spinless) particles obeying Maxwell--Boltzmann statistics. The above
results are extended to quantum systems of many bosons or fermions, as in paper \cite{GP11}.

It was considered some approaches to the description of the evolution of states of quantum
many-particle systems employing the possible modifications of the density operator which
kernel is known as a density matrix. One of these approaches is allowed to describe the
evolution of quantum systems of both finite and infinite average number of particles through
the reduced density operator \eqref{RozvBBGKY} or reduced correlation operators \eqref{sss}
which are governed by the dynamics of correlations \eqref{ghs}.

Above it was established that the notion of cumulants \eqref{cumulantP} of groups of operators
\eqref{rozv_fon-N} underlies non perturbative expansions of solutions for the fundamental
evolution equations, namely for the von Neumann hierarchy \eqref{vNh} of correlation operators,
for the BBGKY hierarchy \eqref{BBGKY} of reduced density operators and for the nonlinear BBGKY
hierarchy \eqref{gBigfromDFBa} of reduced correlation operators, as well as it underlies the
kinetic description of the evolution of states \eqref{f}.

We emphasize that the structure of expansions for correlation operators \eqref{rozvNF-Nclusters},
in which the generating operators are corresponding order cumulant \eqref{cumulantP} of the
groups of operators \eqref{rozv_fon-N}, induces the cumulant structure of series expansions
for reduced density operators \eqref{RozvBBGKY}, reduced correlation operators \eqref{sss}
and marginal correlation functionals \eqref{f}. Thus, in fact, the dynamics of systems of
infinitely many particles is generated by the dynamics of correlations.

The origin of the microscopic description of the collective behavior of quantum many-particle
systems by a one-particle correlation operator that is governed by the generalized quantum
kinetic equation \eqref{gkec} was also considered. One of the advantages of such an approach
to the derivation of kinetic equations from underlying dynamics consists of an opportunity to
construct the kinetic equations with initial correlations, which makes it possible to describe
the propagation of initial correlations in the scaling limits \cite{GTsm},\cite{G14}. In addition,
it was established that in particular case of a mean field approximation for initial states
specified by a one-particle correlation operator the asymptotic behavior of the constructed
reduced correlation operators \eqref{mcc} is governed by the quantum Vlasov kinetic equation \eqref{Vlasov1}.

\end{document}